\documentclass{emulateapj}

\usepackage{graphicx}
\usepackage{psfig}
\usepackage{amsmath}

\def \ergsec{\hbox{erg s$^{-1}$}}
\def \kevcmsq{\hbox{keV cm$^{2}$}}
\def \arcmin {\hbox{$^\prime$}}
\def \arcsec {\hbox{$^{\prime\prime}$}}

\def\spose#1{\hbox to 0pt{#1\hss}}
\def\ltsim{$\mathrel{\spose{\lower 3pt\hbox{$\sim$}}
        \raise 2.0pt\hbox{$<$}}$\thinspace}
\def\gtsim{$\mathrel{\spose{\lower 3pt\hbox{$\sim$}}
        \raise 2.0pt\hbox{$>$}}$\thinspace}
\def \msun {${\rm M_\odot}$}
\def \msunyr{\hbox{{$M_{\odot}$ yr$^{-1}$}}}
\def \mdot{\hbox{{$\dot{M}$}}}

\newcommand\solar{\hbox{{$Z_{\odot}$}}}

\newcommand\tvir{{\hbox{$T_{\rm vir}$}}}
\newcommand{\source}{\mbox{AWM\,4}}

\newcommand\ergs{{\rm erg s$^{-1}$}}

\newcommand{\chandra }{{\em Chandra}}

\newcommand{\lxunits}{\mbox{ergs s$^{-1}$}}
\newcommand{\xmm }{{\em XMM}}

\newcommand\omegam{\hbox{{$\Omega_{\rm m}$}}}
\newcommand\omegalambda{\hbox{{$\Omega_{\Lambda}$}}}
\newcommand\kmsmpc{{\rm km s$^{-1}$ Mpc$^{-1}$}}
\newcommand\ho{\hbox{{$H_{0}$}}}
\slugcomment{Accepted for publication in the Astrophysical Journal Letters}
\shorttitle{The puzzle of the core of AWM 4}
\shortauthors{Gastaldello et~al.}

\begin{document}

\title{Trouble for AGN Feedback ?\\ The puzzle of the core of the Galaxy Cluster AWM 4}

\author {Fabio Gastaldello\altaffilmark{1,2},
         David A. Buote\altaffilmark{1},
         Fabrizio Brighenti\altaffilmark{2,3} 
         \& William G. Mathews\altaffilmark{3},
}
\altaffiltext{1}{Department of Physics and Astronomy, University of
California at Irvine, 4129
Frederick Reines Hall, Irvine, CA 92697-4575}
\altaffiltext{2}{Dipartimento di Astronomia, Universit\`a di Bologna, via
Ranzani 1, Bologna 40127, Italy}
\altaffiltext{3}{UCO/Lick Observatory, University of California at Santa Cruz,
 1156 High Street, Santa Cruz, CA 95064}
\begin{abstract}
The core of the relaxed cluster AWM 4 is characterized by a unique
combination of properties which defy a popular scenario for AGN
heating of cluster cores. A flat inner temperature profile is indicative of a
past, major heating episode which completely erased the cool core, as
testified by the high central cooling time (\gtsim 3 Gyr) and by the
high central entropy level ($\sim 60$ keV cm$^{2}$). Yet the presence of 
a 1.4 GHz active central radio galaxy with extended radio lobes out to 
100 kpc, reveals recent feeding of the central massive black hole. 
A system like AWM 4 should have no radio emission at all if only feedback 
from the cooling hot gas regulates the AGN activity.
\end{abstract}


%%%%%%%%%%%%%%%%%%%%%%%%
\section{Introduction} 
\label{Introduction} 
%%%%%%%%%%%%%%%%%%%%%%%%

The cores of galaxy clusters have been a long standing puzzle since the
early X-ray observations. Initially gas in the cores of clusters was
thought to cool, condense, and flow toward the center, as long as the cooling
time was less than the age of the universe. Most ($\sim$2/3) galaxy clusters
in the local universe satisfy
this condition, with early estimates of their accretion rates as high as
$10^2-10^3$ \msunyr. The problem with this interpretation
was that the mass sink for all this supposedly cooling and condensing gas
has never been found \citep[the classical cooling flow problem or ``mass-sink 
problem'', e.g.][]{donahue04}. X-ray observations with \chandra\ and \xmm\ have
established that there is little evidence for emission from gas
cooling below $\sim$\tvir/3 \citep[the new ``spectroscopic $\dot{M}$ problem'',][]{donahue04}. Just when gas should be cooling most rapidly it
appears not to be cooling at all \citep[e.g,][]{peterson06}.

A compensating heat source must therefore resupply the radiative losses
and many possibilities have been proposed, including thermal
conduction \citep[e.g.,][]{narayan01}, energy released by mergers
\citep[e.g.,][]{motl04} and heating by an active AGN \citep[e.g.,][]{tabor93}.
AGN feedback heating has become the most appealing solution to the
problem \citep[for a recent and detailed review see][]{agnheatingaraa} for the 
following reasons:
1) the high incidence of clusters with short central cooling times
means they cannot be a transient phenomenon, requiring a heating rate
which closely matches the cooling rate. This is difficult to explain, unless
heating rates are coupled to cooling rates. AGN feedback is a natural vehicle
to provide the coupling if the AGN is fed by cooled or cooling gas
\citep{churazov02};
2) there is clear observational evidence for AGN heating in cool core
clusters, since a majority ($\sim$71\%) harbour radio
sources \citep{burns90}. Following the launch of \chandra, ``cavities''
in the ICM of a similar fraction of cool core clusters
\citep[e.g.,][]{dunn06a} have been found which correlate
with the radio emission of the central AGN. These have been
interpreted as bubbles of relativistic plasma inflated by the radio jets
causing $PdV$ heating \citep{churazov02}. Together with weak shocks
associated with the outburst, long
expected in models of jet-fed radio lobes \citep[][]{scheuer74}
and finally detected in deep \chandra\ observations, as for example, in
M87 \citep{forman05}, Hydra A \citep{nulsen05}
and MS 0735+7241 \citep{mcnamara05},
the energies provided by the AGN are not only comparable to those needed
to stop gas from cooling, but the mean powers of the outbursts are well
correlated with the powers radiated \citep{nulsen06};
3) AGN feedback has broader astrophysical implications
for galaxy formation, explaining the truncation of the high end of the galaxy
luminosity function and the symbiosis of black holes and spheroids, forming
an impressive coherent picture. But we are far from a clear understanding,
since ``some loose ends remain'' \citep{binney05}.

Entropy can offer more direct insights into the processes that add and remove
thermal energy from the gas. Motivated by a study of entropy profiles in a
sample of cool core clusters \citep{donahue06}, \citet{voit05c} proposed a
framework for AGN heating in line with the features discussed above and able
to explain the observed core entropy profiles with episodic outbursts of
$\sim 10^{45}$ \ergs\ occurring on a $\sim 10^{8}$ yr timescale. Stronger
outbursts like those seen in Hydra A and MS 0735+7241, which are rarer and
longer lasting may also play a role in extending the time between
outbursts and in elevating the intra-cluster entropy beyond the core region
of clusters. Another clue comes from the \chandra\ observations of two
radio-quiet \citep[at 1.4 GHz in the NVSS,][]{condon98} clusters by \citet{donahue05}: A 1650 and A 2240. Both 
clusters were classified as
strong cooling flows by \citet{peres98}, but their central cooling times
($\sim10^9$ yr) are much longer than the central cooling times of cooling
flow clusters with current radio activity ($\sim10^8$ yr); they also have 
flatter core entropy profiles with larger values of central entropy
(30-40 keV cm$^{2}$) compared to ones ($\sim 10$ keV cm$^{2}$) of the sample 
with active radio central galaxies. 
Either the gas in these clusters experienced such a dramatic episode
of feedback heating sometime in the past that it has not required any
additional feedback for $10^9$ yr or mergers have kept the gas from
cooling and condensing appreciably since the central galaxies were
formed, as for example shown in the simulations by \citet{burns07}.
The fact that
signs of AGN feedback appear only in those cool cores that currently
require feedback has been interpreted as strengthening the case for AGN
heating. In this scenario there is a clear dichotomy between active and
radio quiet clusters: one would expect the cluster population to bifurcate
into systems with strong temperature gradients and feedback and those without
either \citep{donahue05}.

A different scenario for another radio quiet cluster, A 644, in the sample of 
\citet{peres98} was proposed by \citet{buot05a}. In that paper a merger 
scenario was proposed to explain the properties of the
radio-quiet clusters A 644 and A 2589, which have almost isothermal 
temperature profiles and abundance profiles that decrease at large radii. 
\citet{buot05a} suggested that the
center offsets observed in these clusters ($\approx$60 kpc A 644; 
$\approx$10 kpc A2589) indicated that they are
settling down from a past merger event. The larger offset and central 
(\ltsim 50 kpc) disturbance in A644
indicates it is at an earlier stage after the merging. The cluster A2029 
could represent an interesting later stage for these objects. Since A2029 has 
a cool core, a very small center offset ($\approx$4 kpc, smaller than A644 
and A2589), with a WAT, yet its image morphology is otherwise highly relaxed, 
cooling has advanced further and the feedback episode has just started.
It is interesting to note that of the objects in \citet{peres98} as listed in 
\citet{donahue05} lacking 
either an emission line nebula or a strong radio source, two without 
emission lines and radio activity 
\citep[A 644 and A 1689, for the latter see e.g.,][]{girardi97} and one 
with  emission lines but a weak radio source \citep[A 2142,][]{mark00} show 
signs of merger activity.

In this paper we present the properties of the low-mass cluster AWM4 at 
$z$=0.0317 which is a clear counter-example to the dichotomy proposed by 
\citet{donahue05}.
All distance-dependent quantities have been computed assuming \ho = 70 \kmsmpc,
\omegam = 0.3 and \omegalambda = 0.7. At the redshift of $z=0.0317$ 1\arcmin\
corresponds to 38 kpc. All the errors quoted are at the 68\% confidence
limit.

%%%%%%%%%%%%%%%%%%%%%%%%%%%%%%%%
\section{The properties of \source}
\label{x-ray}
%%%%%%%%%%%%%%%%%%%%%%%%%%%%%%%%

%%%%%%%%%%%%%%%%%%%%%%%%%%%%%%%%%%%%%%%%%%%%%%%%%%%%%%%%%%%%%%%%%%%%%%%%%%%%%%
\begin{figure}[th]
\centerline{\psfig{figure=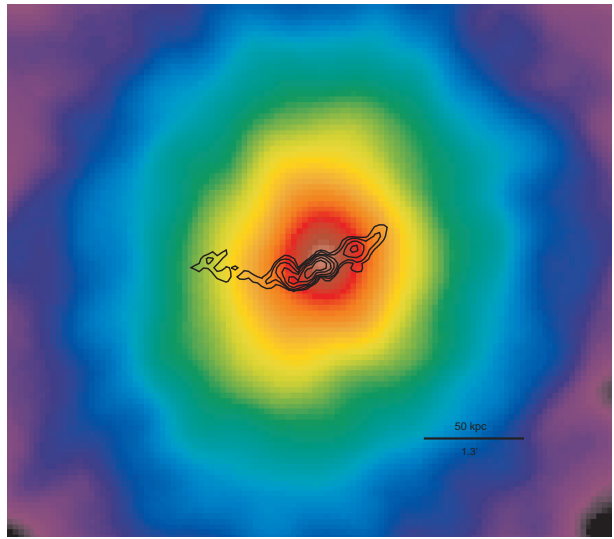,height=0.3\textheight}}
\caption{\label{fig.source} \footnotesize \xmm\ 0.5-2 keV X-ray image of the
core of the cluster AWM 4. The image combines data from the two MOS cameras:
the image has been processed to remove point sources, flat fielded with a
1.25 keV exposure map and smoothed with a 16\arcsec\ gaussian. Radio contours
taken from the VLA FIRST 20cm survey are superimposed in black.
%\vspace{-0.6cm}
}
\end{figure}

%%%%%%%%%%%%%%%%%%%%%%%%%%%%%%%%%%%%%%%%%%%%%%%%%%%%%%%%%%%%%%%%%%%%%%%%%%%%%%

%%%%%%%%%%%%%%%%%%%%%%%%%%%%%%%%%%%%%%%%%%%%%%%%%%%%%%%%%%%%%%%%%%%%%%%%%%%%%%%%
\begin{figure*}[th]
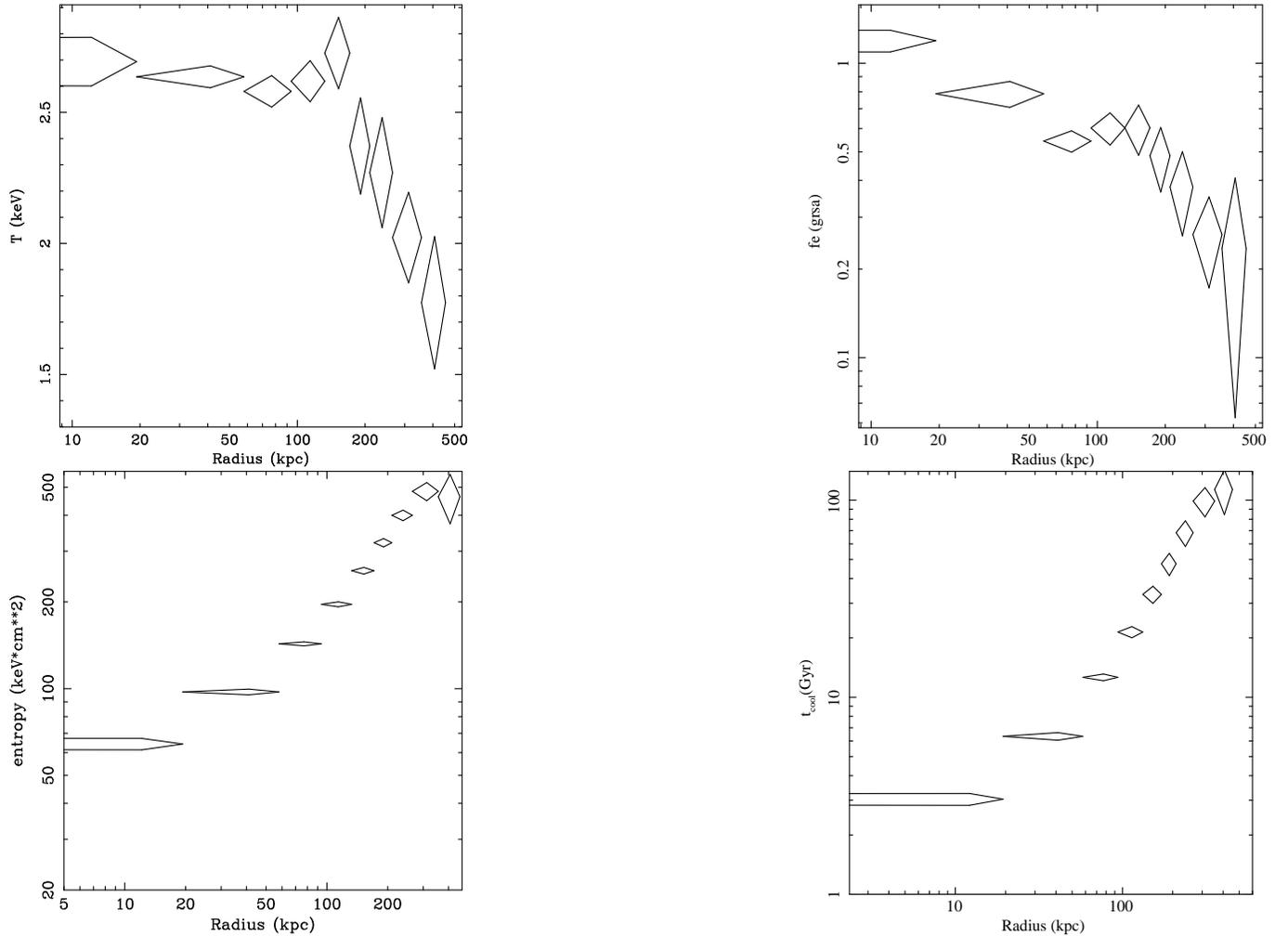

%\vspace{-0.5cm}
\parbox{0.5\textwidth}{
\psfig{figure=awm4_temp_proposal.ps,angle=-90,height=0.28\textheight}}
\parbox{0.5\textwidth}{
\psfig{figure=awm4_fe.ps,angle=-90,height=0.28\textheight}}
\parbox{0.5\textwidth}{
\psfig{figure=awm4_entropy_proposal.ps,angle=-90,height=0.28\textheight}}
\hspace{4.6cm}
\parbox{0.5\textwidth}{
\psfig{figure=cooling_time.ps,angle=-90,height=0.28\textheight}}
\caption{\label{fig.profiles} %\footnotesize
From left to right: the temperature profile of AWM 4 as obtained by the \xmm\
data, the abundance profile, the entropy profile and the cooling time profile.
\vspace{-0.25cm}
}
\end{figure*}
%%%%%%%%%%%%%%%%%%%%%%%%%%%%%%%%%%%%%%%%%%%%%%%%%%%%%%%%%%%%%%%%%%%%%%%%%%%%%%%%

\source\ is a poor cluster whose X-ray emission is extended and
regular (see Fig.\ref{fig.source}) and the peak of the X-ray emission is
coincident with the dominant giant elliptical NGC 6051. It has a bolometric 
luminosity (0.1-100 keV) of $3.93\pm0.06\times10^{43}$ \ergsec\ and a mass 
weighted temperature of $2.48\pm0.06$ keV within 455 kpc, and a mass of 
$1.04\pm0.10 \times 10^{14}$\msun\ within $r_{500} = 708\pm23$ kpc.
It therefore lies at the low end of the mass and 
temperature range defining clusters of galaxies. 28 galaxy members
have been identified, most are absorption line systems with a strong
concentration of early-type galaxies in the center and with a smooth gaussian
velocity distribution centered at the velocity of NGC 6051; 
the velocity dispersion of the system is 
$439^{+93}_{-58}$ km s$^{-1}$ \citep{koranyi02}.
NGC 6051 is considerably brighter than the galaxies around it and recent 
deep $R$-band imaging (Zibetti et al., in preparation) has  
established the fossil nature of this system, according to the definition of 
\citet{jones03}. The \xmm\ data (see Fig.\ref{fig.profiles}, 
upper right panel) show a clear abundance gradient, from $\sim 0.2$ \solar\ 
\citep[in the units of][]{grsa98} at 400 kpc to 
$1.2\pm0.1$ in the inner 20 kpc, another indication of 
a fairly relaxed system \citep[e.g.,][]{degr04a}.
NGC 6051 harbors a powerful radio source with 1.4 GHz flux of $607\pm22$ mJy 
\citep[from the NVSS catalog,][]{condon98} and spectral index $-0.72\pm0.04$ 
\citep[calculated using radio fluxes from 26.3 MHz to 
10.55 GHz taken from NED and from][]{neumann94} corresponding to a total radio 
luminosity of $8.9\times10^{40}$ \ergs\ in the energy band 10 MHz - 10 GHz. 
Radio contours taken from the FIRST 
1.4 GHz radio image \citep{first} are also shown in Fig.\ref{fig.source}.

Given its relaxed appearance both in the X-ray and in the optical 
we included the object in our sample of 16 bright relaxed groups/poor clusters 
observed by either \chandra\ or \xmm\ and chosen to be among the candidates 
best-suited for the application of the hydrostatic equilibrium approximation 
to measure their mass profiles \citep{gasta07a}, where the details of the 
analysis of the available \xmm\ 
observation of \source\ can be found. The \xmm\ data show a unique 
temperature profile (see Fig.\ref{fig.profiles}, upper left panel) for a 
relaxed object, with 
an isothermal core out to 200 kpc \citep[as found also in][]{osul05} and then 
a decline at large radii. 

To characterize quantitatively the entropy profile, determined by computing the
adiabatic constant $K = kTn_e^{-2/3}$ (see Fig.\ref{fig.profiles}, lower left 
panel), we follow \citet{donahue06} and 
fit both a simple power law of the form $K = K_{100}\,(r/100\,\rm{kpc})^{\alpha}$
and the same power law plus a central constant $K = K_0 + K_{100}\,(r/100\,\rm{kpc})^{\alpha}$. We find $K_{100} = 184 \pm 25$ \kevcmsq\ and $\alpha = 0.71\pm0.03$ with $\chi^{2}$/dof = 107/7 for a single power law fit and 
$K_0 = 52 \pm 6$ \kevcmsq, $K_{100} = 122 \pm 51$ \kevcmsq\ and $\alpha = 1.14\pm0.08$ with $\chi^{2}$/dof = 7/6 for a power law plus constant fit. The 
elevated central entropy of \source\ is reflected in the long 
central cooling time \citep[calculated as $t_{cool} = 5/2\,kT/n\Lambda$, e.g.,][see lower right panel of Fig.\ref{fig.profiles}]{peterson06}, $3.0\pm0.2$ Gyr in the inner bin. In this regard \source\ 
shares the same characteristics of the two radio-quiet clusters, A 1650 and 
A 2244, of \citet{donahue05} and would be consistent with the hypothesis 
that a major past AGN outburst has completely erased the cool core. 
But the current 1.4 GHz radio activity implies a recent (not much more than 
$\sim 10^8$ yr ago, the upper limit to the characteristic lifetime, i.e. 
synchrotron age, of the radio source) feeding 
of the central black hole, which is unlikely to come from the ambient hot gas
which has an order of magnitude longer cooling time.

\begin{figure}[t]
\centerline{\psfig{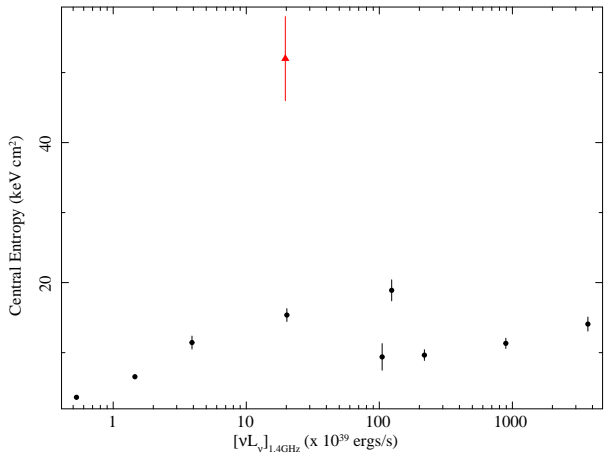}}
\caption{\label{fig.sradio} \footnotesize Central entropy values taken from 
the sample of \citet{donahue06} (black circles) and for \source\ (red triangle). The radio power $\nu L_{\nu}$ at 20 cm is taken from the NVSS \citep{condon98}.
%\vspace{-0.6cm}
}
\end{figure}

The anomalous nature of \source, compared to the objects in the sample of \citet{donahue06}, stands out 
clearly in a plot of central 
entropy versus 20 cm radio power (calculated as $\nu L_{\nu}$ where 
$L_{\nu} = 4 \pi D_{L}^{2} S_{\nu}$, neglecting the K-correction term), as it can be seen in 
Fig.\ref{fig.sradio}. We did not plot the two radio quiet objects of \citet{donahue05} 
because a homogeneous comparison with data at 1.4 GHz is needed: for example a central radio 
source has been detected in A 1650 with a sensitive observation at 327 MHz by \citet{markovic04}, 
but this can be explained as the old remnant of the previous very active AGN. 
On the contrary the \xmm\ data for AWM 4 suggests a surprising lack of low entropy gas 
surrounding an active and bright radio source 
at 1.4 GHz, although they cannot rule out its presence at scales smaller than the inner 20 kpc.
For the clusters in the \citet{donahue06} sample, even though the entropy profiles are more
detailed because of the higher spatial resolution of \chandra, they still rely on temperature 
determinations in inner bins with widths ranging from 7 kpc (A 262, z=0.0163) to 32 kpc 
(PKS 0745-191, z=0.1028) with 4 objects out of 9 with inner bins with width $\gtrsim 20$ kpc.
%
%%%%%%%%%%%%%%%%%%%%%%%%%%%%%%%%%%%%%%%%%%%%%%%%%%%%%%%%%%%%%%%%%%%%%%%%%%%%%%
%
%=====================================================================
% DISCUSSION  DISCUSSION  DISCUSSION  DISCUSSION  DISCUSSION   
%=====================================================================
\section{Discussion}
\label{discussion}

%%%%%%%%%%%%%%%%%%%%%%%%%%%%%%%%%%%%%%%%%%%%%%%%%%%%%%%%%%%%%%%%%%%%%%%%%%%%%%
%%%%%%%%%%%%%%%%%%%%%%%%%%%%%%%%%%%%%%%%%%%%%%%%%%%%%%%%%%%%%%%%%%%%%%%%%%%%%%

Our investigation of current \xmm\ data shows no statistically significant 
surface brightness or spectral variations in the core 
at a level greater than $1-2\sigma$ \citep[as it is also the case for the 
temperature variations quoted in][see their table 5]{osul05}.
In this section we will discuss possible 
scenarios to explain the characteristics of this peculiar source.

There is a clear observational dichotomy between cool core (CC), relaxed clusters
and non-cool core (NCC), disturbed clusters. This is not only based
on the central temperature gradient or central cooling time, but also
for example on the degree of regularity of morphologies 
\citep[as measured by power ratios, which strongly correlates with the strength of the cooling, as 
measured for local and moderate redshift clusters:][]{buot96,bauer05} 
or the presence/absence of a central abundance gradient \citep{degr01}.
A possible new class of objects can include systems like A 1650 and A 2244, 
which have all the characteristics of relaxed objects but lack a cool core 
due to a strong episode of AGN feedback.
Regardless of this possible last class, it is commonly assumed that mergers can transform 
CC clusters into NCC ones, depending on factors such as the initial strength of the 
cooling core, how off-center the merger is and the mass-ratio of the merging
system \citep[][]{mcglyn84,gomez02,ritchie02}. There is still debate and controversial
results from simulations regarding the effectiveness of mergers in 
disrupting cool cores and being responsible for the CC/NCC dichotomy 
\citep{poole06,burns07,mccarthy07}. The inability of current hydro-dynamical simulations in 
reproducing realistic cool cores may be one of the reason for these controversial results.
Nevertheless, it is crucial to discuss a merger hypothesis for the properties 
of \source. The central radio source has some similarities with a 
wide-angle-tail (WAT) source, which it has been argued can be shaped by ICM
ram pressure produced in cluster mergers \citep[e.g.,][]{burns94,gomez97}.
This is a well motivated hypothesis for some of the most relevant cases 
like the ones presented in \citet{gomez97} which present clear evidence of
disturbed X-ray morphologies, but it can not hold in cases like the one of 
\source. Small scale disturbances due to accretion of small mass 
objects, like the simulations described in \citet{ascasibar06} can 
reproduce the characteristic X-ray spiral structure seen for example in the 
core of Perseus and provide the necessary relative ICM motions
to explain the WAT-like jet bending. Alternatively jet bending can result from 
instabilities, preexisting radio plasma, as discussed for Hydra A 
\citep{nulsen05}, or by jet precession, as in the spectacular case 
of the cluster RBS797 \citep{gitti06}. 
A good comparison can be made with the recent \chandra\ observation of the 
prototypical WAT host cluster A 1446 \citep{douglass07}: 
in this cluster there is a complex X-ray morphology in the 
inner 200 kpc and there is a flat abundance profile, a sensitive indicator of 
an ongoing merger. \source\ has very different properties in the X-rays and 
in the optical, in particular its fossil property would be difficult to 
understand in a recent major merger scenario. The same conclusion has 
been reached by \citet{osul05}. 

Clearly given the spatial resolution of the \xmm\ data, the possibility of 
the accretion of a low-mass object, like a small group, a cool galaxy corona,
able to survive stripping and thermal evaporation \citep{sun07},
or even a gas rich dwarf galaxy, can not be ruled out as the cause of cooled 
gas that triggered the radio jet in \source.

It is also useful to compare \source\ with the suggested
evolutionary scenario discussed in \citet{buot05a}.
Systems like A 644 are just settling down from a major merger to establish 
the initial cooling/feedback loop and there is no radio emission  until the
core is basically relaxed.
A possible interpretation is that \source\ would instead have represented the 
final stage of the cycle of AGN feedback, like A 1650 and A 2244, if it were 
not for a small-merger/accretion related event. The limits posed by the 
\xmm\ image to the center offset is $8\pm2$ kpc, but this determination is 
hampered by low \xmm\ spatial resolution (the PSF FWHM on-axis 
corresponds to 2 kpc at the redshift of the source). 

It is still possible to reason along the lines of the model for example 
of \citet{voit05c} but we have to allow the previous AGN outflow of 
$\sim 10^{46}$ \lxunits (capable of raising the central entropy to the 
observed $\sim50$ \kevcmsq) of having been not able to completely stop 
cooling near the very core, much alike the cool coronae observed 
in NGC 3842 (in A 1367) and NGC 4874 (in Coma) by \citet{sun05}. In these 
sources if the AGN mechanical power, associated with the observed radio 
emission, has the same magnitude observed in cool core clusters, it would 
have completely disrupted the observed X-ray soft coronae. Remaining cooling 
gas (or gas supplied by stellar mass loss) could have triggered the current 
outburst: assuming the energy released 
by accretion onto a black hole to be $\eta$\mdot$c^2$ with $\eta=0.1$ we have 
$\mdot=0.0017\,L_{X,43}$ \msunyr, where $L_{X,43}=10^{43}$ \ergsec; an accretion 
rate as low as 0.017 \msunyr\ could sustain a total mechanical luminosity 
of $10^{44}$ \ergsec, a thousand times higher than the radio luminosity. 

Measurements of X-ray cavity sizes and surrounding gas pressure have provided unique estimates of 
the ratio of jet power to synchrotron power.
The jet power determined from X-ray cavity data shows a clear correlation 
with synchrotron power (core plus lobes); the median ratio 
of jet (cavity) power to synchrotron power is $\sim 100$, with mean $\sim 2800$ owing to large 
(ranging from a few to a few thousand) scatter \citep[][in particular their 
Fig.7]{birzan04,agnheatingaraa}; this is interestingly on the high side of 
theoretical expectations \citep[e.g.,][]{deyoung93}.
Given its current radio luminosity the central 
radio source in AWM 4 can well provide up to $10^{43}-10^{44}$ \ergs\ of 
mechanical energy into the ICM. These luminosities would be sufficient to carve X-ray 
cavities in the ICM: if for example the mechanical luminosity associated with 
the radio jets in AWM4 is $10^{43}$ \ergs\ and assuming an enthalpy of $4pV$, 
appropriate for cavities dominated by relativistic particles, and a duration 
of the outburst of $10^7$ years, the energy
created would result in a pair of spherical cavities with radius $\approx$3.5 kpc located 
within $\approx$20 kpc from the center, perhaps coincident with the inner 
bright radio ``knots'' at 21\arcsec\ (13 kpc) and 34\arcsec\ (21 kpc) from 
the center.
Cavities of these sizes would produce a level of contrast in surface brightness above 30\%, which 
is the contrast of evident bubbles like the ones in A 2052 and Hydra A.
If cavities are detected in \source, it would establish that
AGN mechanical heating has taken place in the past $\sim 10^{8}$ yr, given the
fact that any cavities that might be seen today may only have come from 
the currently visible radio jet: for example the buoyancy time scale 
for a bubble of 3.5 kpc at R=21 kpc is $9\pm2\times10^{7}$ yr, calculating the gravitational 
acceleration as $g=GM(<R)/R^2$ using the mass profile derived in \citet{gasta07a};
using instead $g\sim2\sigma^{2}/\rm{R}$ where $\sigma=243\pm24$km/s 
\citep{davies87} is the stellar velocity dispersion of the central galaxy would result in
$8\pm2\times10^{7}$ yr.

Another indication of the energy of the previous outburst which has 
disrupted the cool core of \source\ comes from the analysis of 
\citet{rebusco06}. By comparing the observed peaked iron abundance profile 
for a sample of groups and clusters, including Perseus, with the prediction 
of a simple model invoking the metal ejection from 
the BCG and the subsequent diffusion of metals by stochastic gas motions, 
they found for \source\ the largest value of the diffusion coefficient, 
implying the strongest level of gas mixing.
%
%=====================================================================
% CONCLUSION  CONCLUSION  CONCLUSION  CONCLUSION  CONCLUSION  
%=====================================================================
\section{Conclusions}
\label{conclusion}

\source\ provides an early critical test to the new born theory of AGN 
feedback, and it exposes our poor 
understanding of the details of the fueling and triggering of AGN 
outbursts, including the part, if any, played by mergers, as for example 
highlighted in the future issues to be addressed by \citet{agnheatingaraa}.
We must understand the new cycle of heating and cooling of 
a relaxed galaxy cluster, 
as depicted for example in the AGN feedback loop by \citet{tucker07}, 
and its connection to the pre-\chandra\ loop of mergers and 
relaxation in cool cores \citep[e.g.,][]{buot02b}. 

%=====================================================================
% ACKNOWLEDGEMENTS  ACKNOWLEDGEMENTS  ACKNOWLEDGEMENTS 
%=====================================================================
\begin{acknowledgements}
We would like to thank S. Ettori, M. Gitti, P.J. Humphrey, D. Pierini,  
S. Zibetti and the anonymous referee for useful comments.
F.G. and D.A.B. gratefully acknowledge partial support from
NASA grant NAG5-13059, issued through the Office of Space Science
Astrophysics Data Program.
This work is based on observations obtained with \xmm\, an ESA science 
mission with
instruments and contributions directly funded by ESA member states and
the USA (NASA).  
\end{acknowledgements}

%=====================================================================
% REFERENCES  REFERENCES  REFERENCES  REFERENCES  REFERENCES  
%=====================================================================

%\newpage

\end{document}